\documentclass[12pt]{article}
\usepackage{times}
\usepackage[letterpaper, portrait, margin=1in]{geometry}
\usepackage{useltp_ksp_1.2}
\usepackage[utf8]{inputenc}
\usepackage{enumitem,amssymb}
\usepackage{graphicx}
\usepackage{ragged2e}
\newlist{thematic}{itemize}{8}
\setlist[thematic]{label=$\square$}
\usepackage{pifont}
\usepackage{setspace,wrapfig}
\usepackage{multicol}
\usepackage[compact]{titlesec}
\newcommand{\cmark}{\ding{51}}%
\newcommand{\done}{\rlap{$\square$}{\raisebox{2pt}{\large\hspace{1pt}\cmark}}%
\hspace{-2.5pt}}

\begin{document}
\begin{flushleft}

\huge
Astro2020 Science White Paper \linebreak

Toward Finding Earth 2.0: Masses and Orbits of Small Planets with 
Extreme Radial Velocity Precision \linebreak
\normalsize

\noindent \textbf{Thematic Areas:} \hspace*{60pt} \done Planetary Systems \hspace*{10pt} $\square$ Star and Planet Formation \hspace*{20pt}\linebreak
$\square$ Formation and Evolution of Compact Objects \hspace*{31pt} $\square$ Cosmology and Fundamental Physics \linebreak
  $\square$  Stars and Stellar Evolution \hspace*{1pt} $\square$ Resolved Stellar Populations and their Environments \hspace*{40pt} \linebreak
  $\square$    Galaxy Evolution   \hspace*{45pt} $\square$             Multi-Messenger Astronomy and Astrophysics \hspace*{65pt} \linebreak
  
\textbf{Principal Author:}

Name:David R. Ciardi	
 \linebreak						
Institution: Caltech/IPAC-NASA Exoplanet Science Institute 
 \linebreak
Email: ciardi@ipac.caltech.edu
 \linebreak
Phone:  +1-626-395-1834
 \linebreak
 
\textbf{Co-authors:} (names and institutions)
Jacob Bean (University of Chicago), Jennifer Burt (MIT), Diana Dragomir (MIT/UNM), Eric Gaidos (University of Hawaii at Manoa),  Marshall C. Johnson (The Ohio State University), Eliza Kempton (University of Maryland), Quinn Konopacky (UC San Diego), Michael Meyer (University of Michigan), Johanna Teske (Carnegie Institution), Lauren Weiss (University of Hawaii at Manoa), George Zhou (CfA-Harvard \& Smithsonian)\\
\vspace{10pt}
\textbf{Co-signers:} (names and institutions)
David Ardila (Jet Propulsion Laboratory), Jaehan Bae (Carnegie Institution), Amy Barr (Planetary Science Institute), Alan Boss (Carnegie Institution), Steve Bryson (NASA Ames Research Center), Derek Buzasi (Florida Gulf Coast University), Jessie Christiansen (Caltech/IPAC-NExScI), Jonathan Crass (University of Notre Dame), Thayne Currie (NASA-Ames Research Center), Drake Deming (University of Maryland), Chuanfei Dong (Princeton University), Courtney Dressing (University of California, Berkeley), Sarah Dodson-Robinson (University of Delaware), Eric Ford (Penn State University), Benjamin Fulton (Caltech/IPAC-NExScI), Scott Gaudi (OSU), Dawn Gelino (Caltech/IPAC-NExScI), Hannah Jang-Condell (University of Wyoming), Stephen R. Kane (UC Riverside), Irina Kitiashvili (NASA Ames Research Center \& BAERI), Laura Kreidberg (CfA - Harvard \& Smithsonian), Rogers Leslie (University of Chicago), Michael Line (Arizona State University), Mercedes L\'{o}pez-Morales (CfA-Harvard \& Smithsonian), Patrick Lowrance (Caltech/IPAC-Spitzer), John Mather (Goddard Space Flight Center), Dimitri Mawet (Caltech/JPL), Eliad Peretz (Goddard Space Flight Center), Peter Plavchan (George Mason University), Tyler Robinson (Northern Arizona University), Joseph E.~Rodriguez (CfA-Harvard \& Smithsonian), Sam W.~M.~Ragland (Keck Observatory), Angelle Tanner (Mississippi State University), Gerard van Belle (Lowell Observatory), Sharon xuesong Wang (Carnegie-DTM) 
\end{flushleft}

\justifying
\textbf{Abstract:}
Having discovered that \emph{Earth-sized} planets are common, we are now embarking on a journey to determine if \emph{Earth-like} planets are also common. Finding Earth-like planets is one of the most compelling endeavors of the 21st century -- leading us toward finally answering the question ``Are we alone?''

To achieve this forward-looking goal, we must determine the masses of the planets; the sizes of the planets, by themselves, are not sufficient for the determination of the bulk and atmospheric compositions.  Masses, coupled with the radii,  are crucial constraints on the bulk composition and interior structure of the planets and the composition of their atmospheres, including the search for biosignatures. Precision radial velocity is the most viable technique for providing essential mass and orbit information for spectroscopy of other Earths. 

The development of high quality precision radial velocity instruments coupled to the building of the large telescope facilities like TMT and GMT or space-based platforms like EarthFinder can enable very high spectral resolution observations with extremely precise radial velocities ($\sim$cm/s) on minute timescales to allow for the modeling and removal of radial velocity jitter.  Over the next decade, the legacy of exoplanet astrophysics can be cemented firmly as part of humankind's quest in finding the next Earth -- but only if we can measure the masses and orbits of Earth-sized planets in habitable zone orbits around Sun-like stars. 

Toward this goal, we have three major recommendations:\\
\noindent 1. We endorse the findings and recommendations published in the National Academy reports on Exoplanet Science Strategy and Astrobiology Strategy for the Search for Life in the Universe. This white paper extends and complements the material presented therein.\\
\noindent 2. Specifically, the US should invest in the EPRV initiative as recommended by the National Academies of Sciences Exoplanet Science Strategy Report \\
\noindent 3. The US should invest in the instrumentation, telescope facilities (ground and space-based) and advanced tools for statistical modeling of stellar variability  necessary to obtain the radial velocity observations at the precision and quality sufficient to determine the masses and orbits of Earth-sized planets in the habitable zones of Sun-like stars.  The most compelling systems will be discovered in both hemispheres, and thus, facilities need to access the whole sky.

\pagebreak
\noindent\underline{\textbf{Finding Earth 2.0:}}
Discovery and characterization of Earth-like planets in the habitable zones of Sun-like stars remains one of the highest level goals of exoplanet research today. We have moved from an era of pure discovery to an era of exoplanetary characterization, and have shown that ``small" planets are common around main-sequence stars and many orbit within the ``habitable zone" of their host stars \cite{thompson2018}.  But transit surveys, like Kepler/K2, TESS and PLATO \cite{howell2014, ricker2015,rauer2016} only provide a planet's radius. Direct imaging surveys from the ground (e.g., 30-meter telescopes) or from space (e.g., HabEx or LUVOIR) will find potentially Earth-like planets, but the planetary masses (and radii) will be unknown.  To fully characterize and understand the composition and atmosphere of the planets, the planet mass must be determined. The NAS Exoplanet Science Strategy report states \emph{``Mass is the most fundamental property of a planet, and knowledge of a planet’s mass (along with a knowledge of its radius) is essential to understand its bulk composition and to interpret spectroscopic features in its atmosphere. If scientists seek to study Earth-like planets orbiting Sun-like stars, they need to push mass measurements to the sensitivity required for such worlds \cite{nas2018}''}.
\begin{figure}[hb]
\vspace{-0pt}
    \begin{minipage}[c]{0.45\textwidth}
        \includegraphics[width=\textwidth]{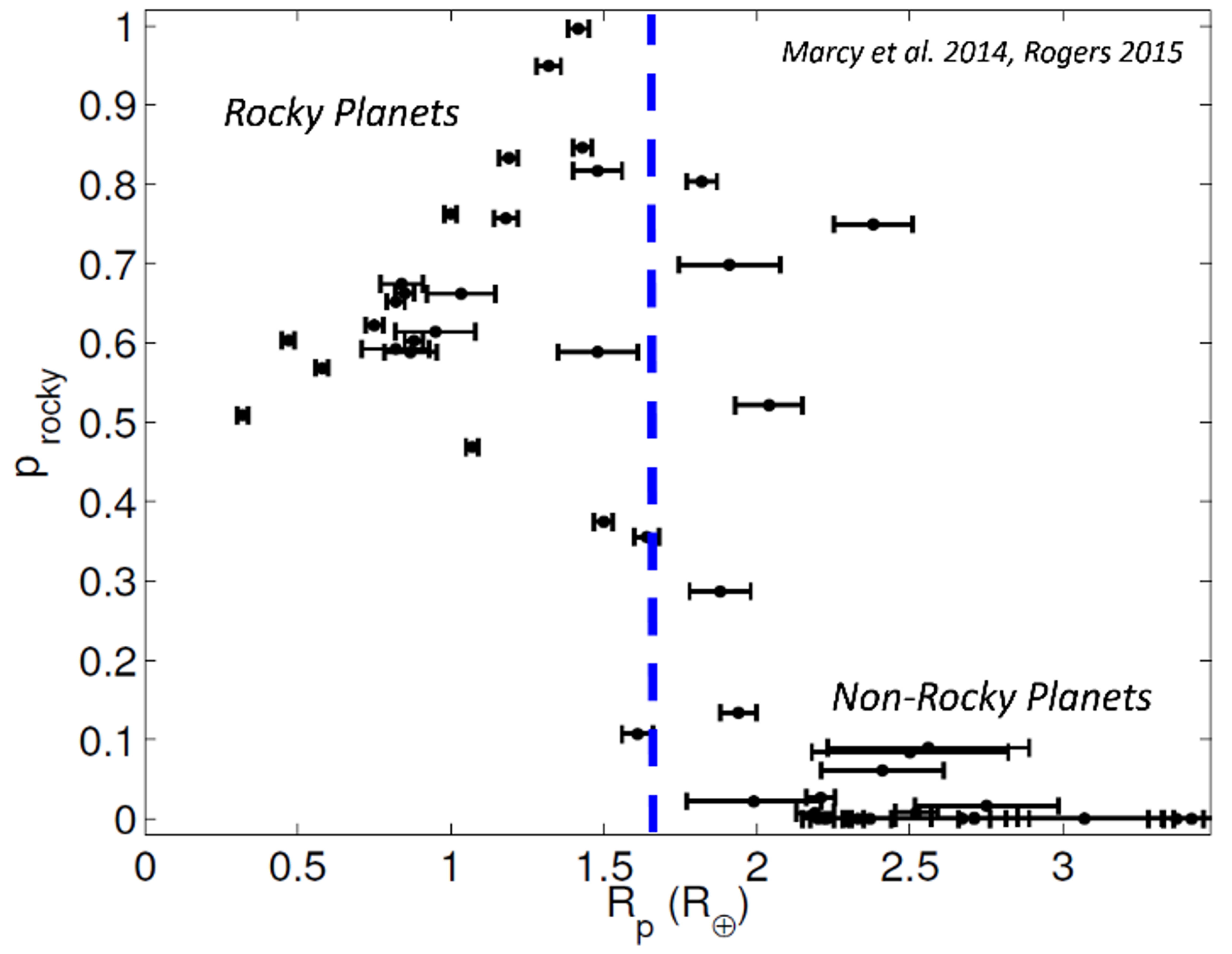}
    \end{minipage}\hfill
    \begin{minipage}[c]{0.52\textwidth}
        \vspace{-30pt}
        \caption{Probability that a planet is rocky as a function of planet radius. The transition from ``non-rocky'' to ``rocky'' planets occurs near a planet radius of $\sim$1.6 R$_\oplus$ and is very sharp – spanning only $\sim0.2$ R$_\oplus$ in radius \cite{marcy2014,rogers2015,fulton2017,wolfgang2016,ning2018}.}
    \end{minipage}
    \vspace{-13pt}
\end{figure}

\noindent\underline{\textbf{Planetary Masses Are Needed For Bulk Compositions: }}
Planetary masses, coupled with radii, provide us with an understanding of the bulk density and the surface gravity. Masses and radii combined can be compared to planetary interior models to place constraints on the bulk composition (silicates, vs. metals vs. low molecular-weight ices vs. non-condensible hydrogen-helium gas), and thus, tell us whether or not planets are rocky and terrestrial or gaseous and Neptunian.  
\begin{figure}[h]
\vspace{-0pt}
    \begin{minipage}[c]{0.45\textwidth}
        \includegraphics[width=\textwidth]{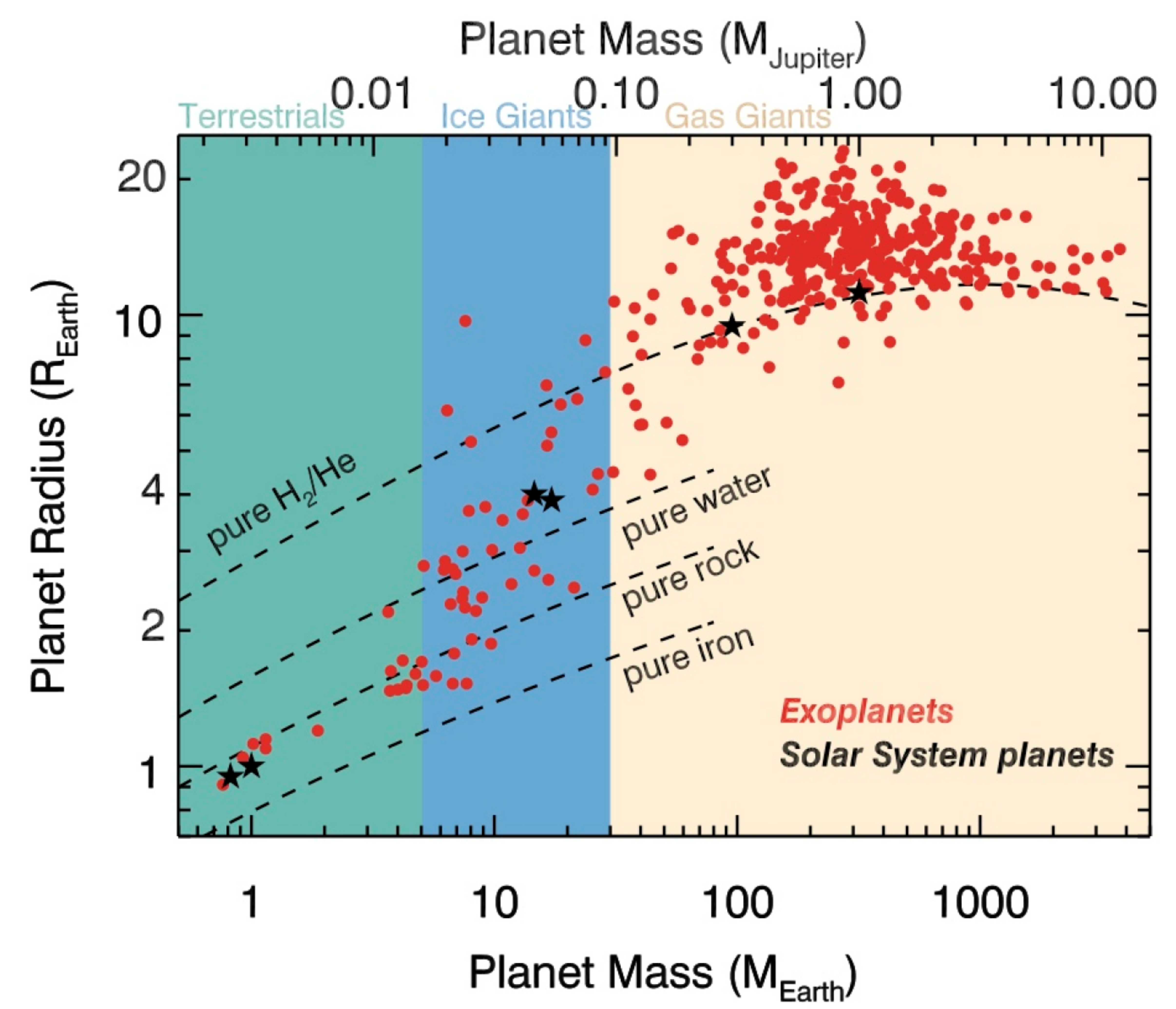}
    \end{minipage}\hfill
    \begin{minipage}[c]{0.5\textwidth}
        \vspace{-5pt}
        \caption{Masses and radii for the 418 exoplanets with fractional measurement errors less than 20 percent (red circles, NASA Exoplanet Archive \cite{akeson2013}). The Solar System planets are indicated by black stars. The dashed lines show the predictions in \cite{fortney2007} of theoretical models for simple compositions.There are only $\sim$dozen planets with measured masses and radii in the terrestrial category because of the difficulty in measuring the masses of these small planets.  Adopted from NAS Exoplanet Strategy Report \cite{nas2018}.}
    \end{minipage}
    \vspace{-13pt}
\end{figure}

Typically, planets with radii $\sim$1.6 R$_\oplus$ or smaller have densities consistent with rocky compositions, while planets with radii $\sim$1.8 R$_\oplus$ or larger planets have densities that are more consistent with a substantial hydrogen-helium envelope around a rocky or icy core (Figures~1 and 2). However, the transition is not uniform nor is it absolute with clear examples of low density small planets and high density large planets.    Our ability to map this transition and the intrinsic scatter within the mass-radius relation is limited by the precision of mass measurements and is currently confined to short orbital period planets.  \textbf{Precise masses of Earth-sized planets in habitable zone orbits are needed to find and understand the frequency and distribution of true Earth-like analogs.}\\

\noindent\underline{\textbf{Planetary Masses Are Needed For Atmospheric Composition: }}
Finding rocky planets is only the first step in finding Earth-like worlds.  Measurements of the atmospheric composition are needed to determine if the planet has an atmosphere, if the planet could support life, and if the planet shows signs of atmosphere biosignatures. Transiting planets provide a unique opportunity to determine the atmospheric composition of the planets via transmission spectroscopy during primary transit and emission spectroscopy during secondary eclipse (see Astro2020 White Papers by Dragomir et al. and Lopez-Morales et al.).  But these techniques require knowledge of the mass in order to determine the compositional abundances of the atmospheres.
\begin{figure}[h]
\vspace{-0pt}
    \begin{minipage}[c]{0.45\textwidth}
        \includegraphics[width=\textwidth]{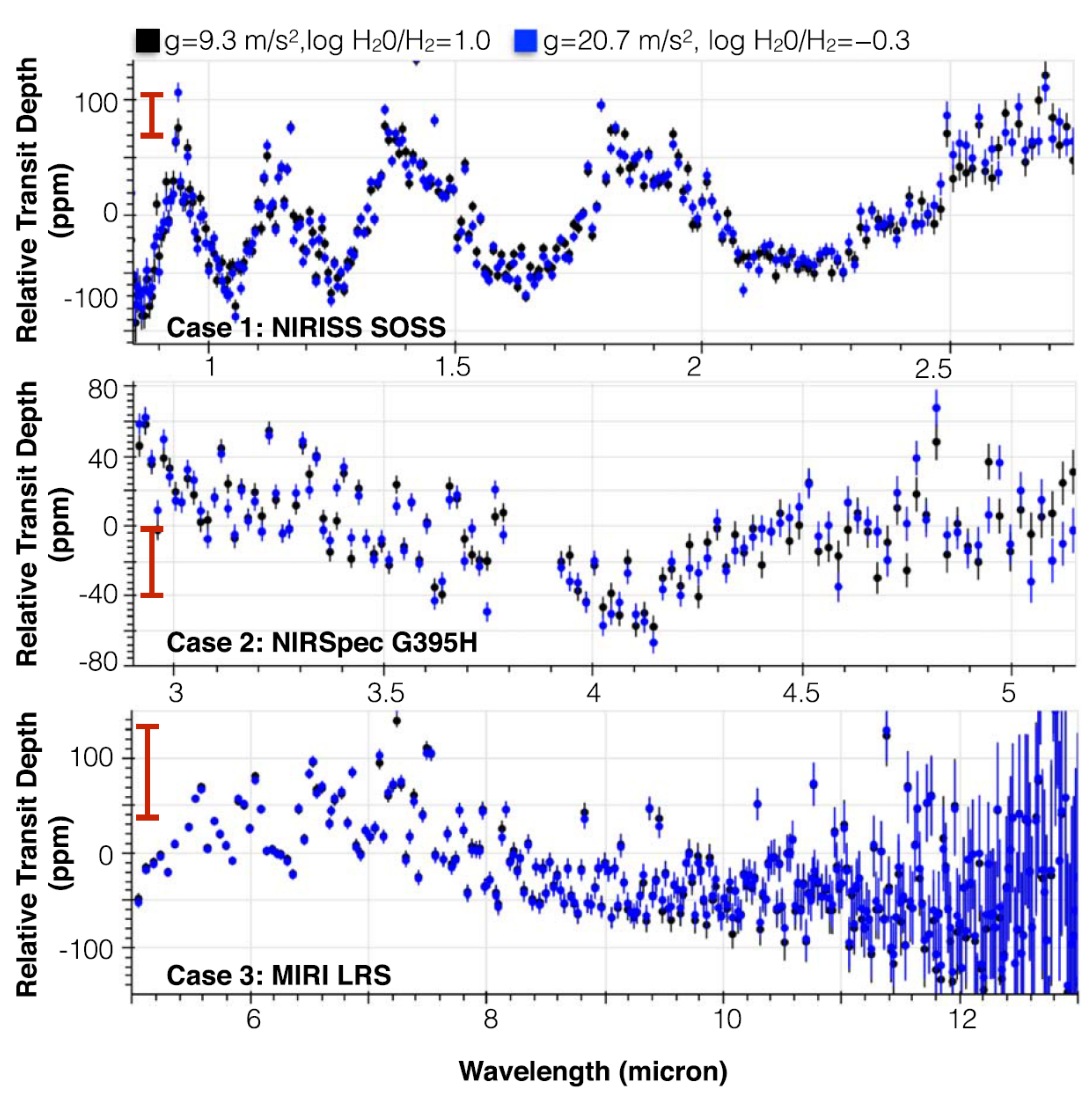}
    \end{minipage}\hfill
    \begin{minipage}[c]{0.5\textwidth}
        \vspace{-20pt}
        \caption{Simulated JWST spectra for a GJ1214-like planet where the mass of the planet is unknown and, therefore, the surface gravity is unknown to within a factor of 2 (g=9.3 m/s$^2$ vs g=20.7 m/s$^2$).  The two different models have abundances differences of more than a factor of 20 and are indistinguishable.  Only with mass measurements can the degeneracy be broken \cite{batalha2017}.}
    \end{minipage}
    \vspace{-13pt}
\end{figure}

The depth of the features observed in the spectra scale directly with the scale height of the atmosphere at those wavelengths:
$H = kT/\mu g_p \propto 1/\mu M_p$
where $T$ is the temperature of the atmosphere, $g_p$ is the surface gravity, $\mu$ is the mean molecular weight, and $M_p$ is the mass of the planet.  The planet mass ($M_p$) and the bulk atmospheric composition ($\mu$) are degenerate with each other; however, if the planet mass can be determined separately, the measured scales heights can be used to determine the atmospheric composition (Fig~3).  More detailed measurements of atmospheric properties (e.g., mixing ratios of molecules found in the atmosphere) depend on first determining the scale height of the atmosphere \cite{sing2016}. Thus, the combination of the planetary mass and radius (i.e., the surface gravity) is required for a determination of the atmospheric structure \cite{kempton2014} and interpreting atmospheric spectra.   \textbf{Masses are needed in order to determine if the discovered Earth-mass planets may indeed be Earth-like in their atmospheric characteristics.}\\

\begin{figure}[h]
    \begin{center}
        \includegraphics[width=0.70\textwidth]{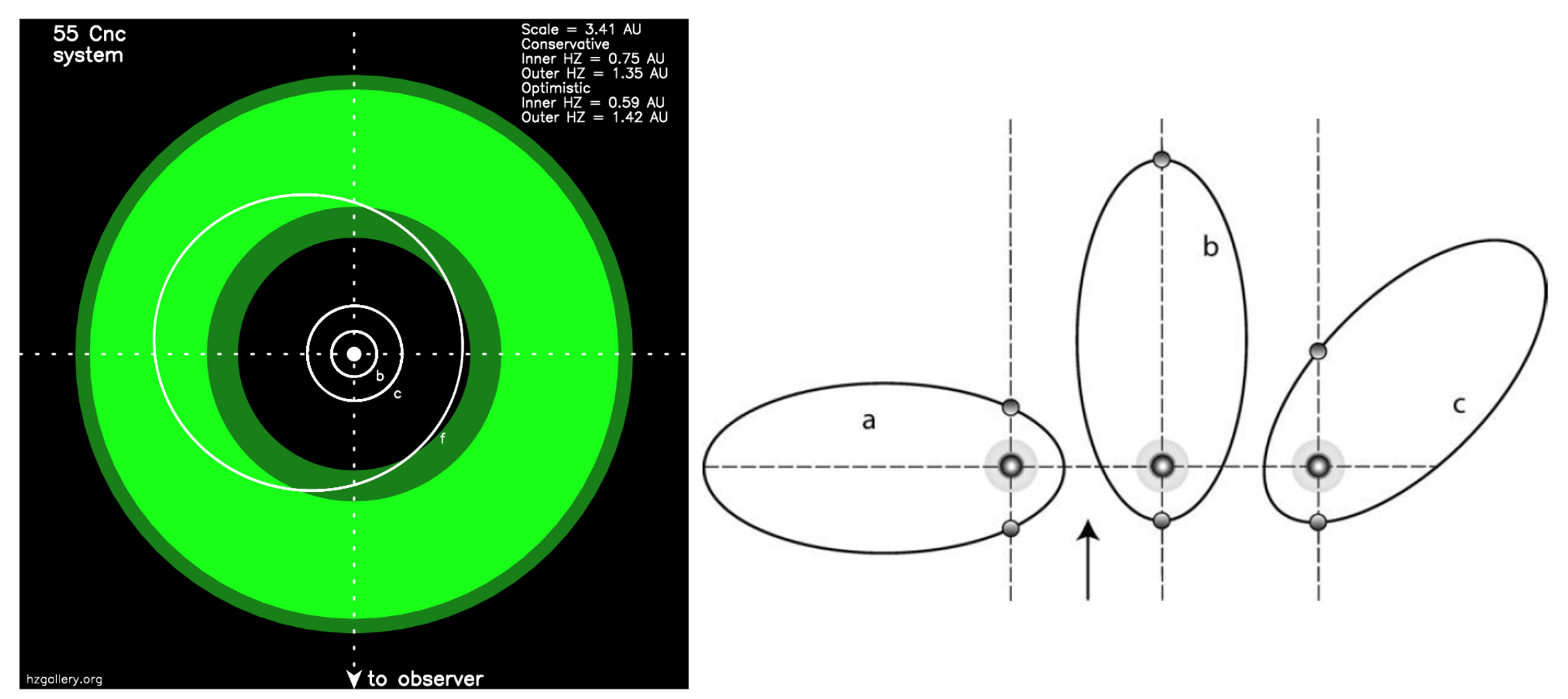}
        \caption{\textit{Left:} Plot of the 55 Cnc system showing that the f planet is on a sufficiently eccentric orbit that the planet enters and exits the Habitable Zone of its star. Figure adopted from the Habitable Zone Gallery \cite{kane2012}. \textit{Right:} Schematic of how the orientation of an eccentric orbit can affect the timing and duration of a secondary eclipse for the same orbital period but oriented differently towards the observer (arrow).  Adopted from \cite{kane2009,beichman2014}.}
    \end{center}
    \vspace{-13pt}
\end{figure}
\noindent\underline{\textbf{Orbit Determinations Are Needed for Demographics and Observing Eclipses:}}
Understanding how unique our Solar System is requires understanding the distribution of planets and their orbital properties.  Transit detection by itself is not sufficient as the transit events reveal little about the eccentricity or obliquity of the orbits. Radial velocity observations used to determine the masses of the planets also reveal the planetary orbits. Clues to the formation and evolutionary history of a planetary system are contained within the current orbital dynamics of the systems.  For example, if a planet is in an eccentric orbit, it may enter and exit the habitable zone of its star, and such planets may not be suitable for life (Fig.~4).  

Further, the dynamical history of the planetary systems is fossilized into the orbital configuration now observed enabling exploration of possible migrations and the determination of the existence of more massive planets further out which may foster the delivery of volatiles into the inner planetary systems. Finally, orbits are not always circular or oriented in a manner that would make the secondary eclipses correctly predicted from just knowing when the primary transits occur.  Orbital determinations are necessary if secondary eclipse observations are to be made - for without knowing the orbit, the timing of the eclipse is unknown (Fig.~4).

Together with measurements of the orbital obliquities of transiting planets (see white paper by Johnson et al.), masses and orbits allow for a full dynamical characterization of planetary systems that may host Earth-like analogs. Radial velocity measurements can find additional non-transiting planets, or long-term trends indicating the presence of long-period planets. All of these aspects help us to achieve a fuller understanding of planetary systems and the context in which the Earth-mass planets reside.

\vspace*{0.1in}
\noindent\underline{\textbf{Extreme Radial Velocity Precision Is Needed for Mass and Orbit Determinations:}} The radial velocity technique, which benefits significantly from the pre-determined discovery and period determination from the transit signatures of the planets, is the most direct way of determining the planetary masses and orbits. Extreme precision radial velocity measurements enabled by the next generation of instruments and the development of the 30-meter telescopes (TMT and GMT) or perhaps a space-based platform (e.g., EarthFinder) will enable mass measurements of Earth-mass planets orbiting within the habitable zones of their stars. 

Measurement of the masses of small planets is currently limited to those systems with the shortest orbital periods (and hence highest irradiances) around the brightest host stars - as those stars have been the practical targets for precision radial velocity followup. As a result, the small planet mass-radius relationship is much less known for planets in wider orbits and lower irradiance, and it's virtually unexplored for planets in orbits within the habitable zone of their host stars. Of the known $\sim4000$ confirmed exoplanets \cite{akeson2013}, there are 672 planets known with orbits of 100 days or longer.  Of these, only 14 planets have both the radii and masses measured. And of these planets, all are Saturn-mass or larger.  However, there are 132 known exoplanets with orbital periods greater than 100 days with measured radii from transits - and of these, 75\% are Neptune-sized or smaller (Fig~5).  \textbf{The planets are there; we need the masses.}\
\begin{figure}[h]
\vspace{-0pt}
    \begin{minipage}[c]{0.425\textwidth}
        \includegraphics[width=\textwidth]{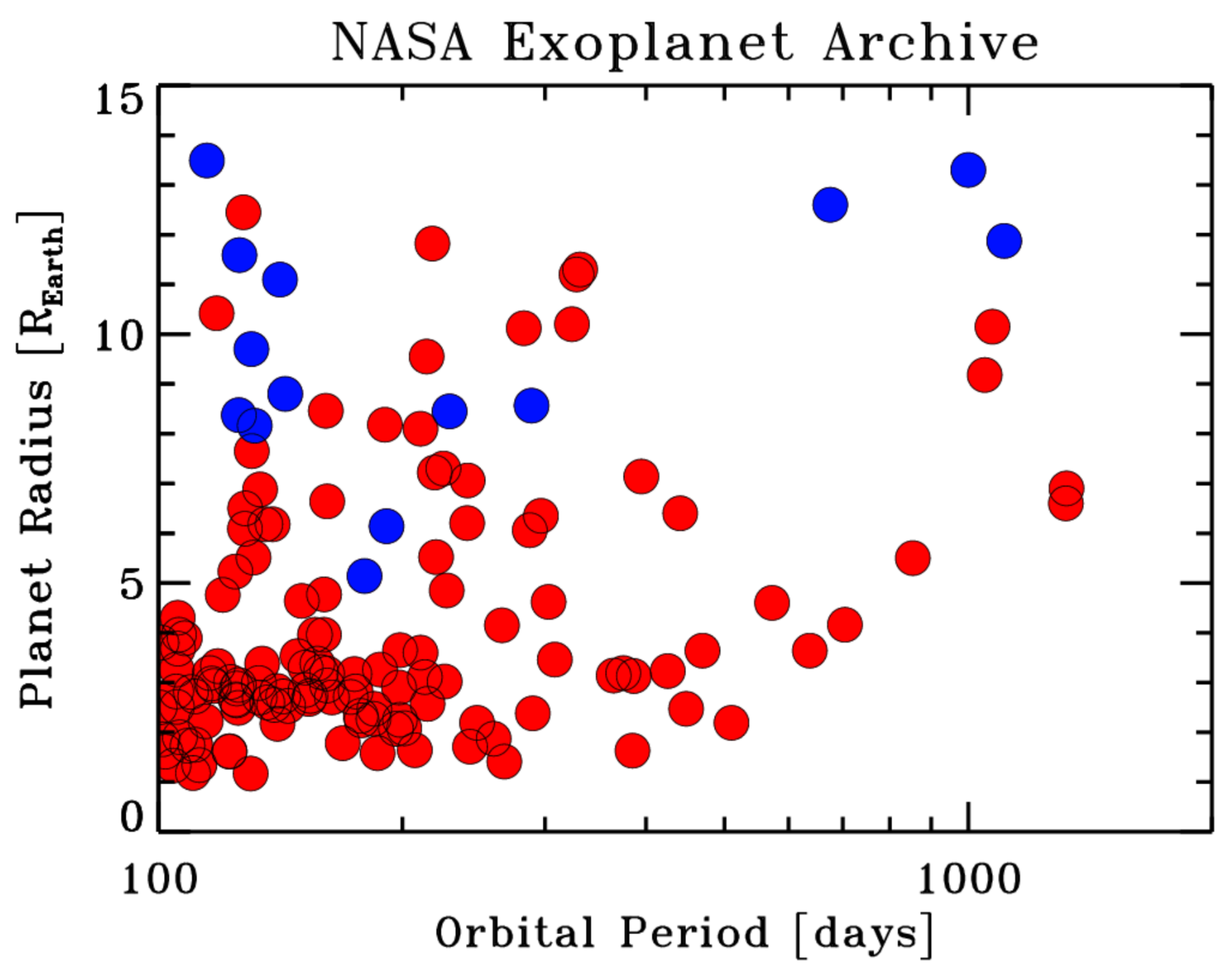}
    \end{minipage}\hfill
    \begin{minipage}[c]{0.55\textwidth}
        \vspace{-20pt}
        \caption{Planet radius vs orbital period plot of the known exoplanets with orbital periods longer than 100 days (red) where planets with mass determinations are highlighted (blue). Even before PLATO, there are $\sim$100 planets identified by Kepler that are Earth-sized and in relatively long orbits that are potentially ripe for mass determinations. Data adopted from the NASA Exoplanet Archive \cite{akeson2013}.}
    \end{minipage}
    \vspace{-13pt}
\end{figure}

The Keplerian radial velocity signal of a Sun-Earth analog is $\sim 9$ cm/s -- below the state of the art in radial velocity precision (0.5-1 m/s).  As the amplitude of the radial velocity signal and the sizes of the habitable zones both scale inversely with the stellar mass, small planets in the habitable zone of low-mass stars ($M_* < 0.3M_{\odot}$) are potentially accessible to today's instrumentation (Fig.~6), but whether these planets -- which are exposed to elevated tides, extreme UV irradiation, and prolonged pre-main sequence irradiation -- are Earth-like remains unclear \cite{scalo2007}.  

The 8-10m class telescopes, if equipped with instrumentation that can reach 1 cm/s, will still play a major scientific role as they can observe the candidates around the brighter stars.  Currently, the majority of the small planets in habitable zone orbits around G and K dwarfs that have been found by Kepler and (eventually by PLATO) orbit stars that are fainter than $V \gtrsim 9$ mag \cite{barclay2018, rauer2016}. Thus, in addition to the precision radial velocity instrument improvements, the development of the large telescopes or a space-based platform will to enable the observation of all the Earth-sized planets. 

Stellar oscillations, granulation phenomena, and chromospheric activity phenomena, which are often collectively referred to as stellar ``jitter'', can in practice limit the obtainable level of radial velocity precision for planet detection. The magnitude of these effects for the Sun are on the order of  1 m/s, with timescales ranging from a few minutes (p-mode oscillations), to tens of hours (granulation phenomena), to tens of days (activity phenomena), and possibly even tens of years (long-term activity cycles). High cadence observations with large telescopes or with a space-based platform can help mitigate the stellar ``jitter'' by sampling the radial velocities at a sufficiently high rate and signal-to-noise to enable modeling and removal of the jitter \cite{lovis2006}, as the data contain sufficient information to recognize the effect of star spots on the apparent Doppler shift \cite{davis2017} and to detect planets \cite{jones2017}.  

In addition to stellar jitter, unresolved spectral lines in the Earth's atmosphere (micro-tellurics) can induce velocity shifts of $\sim 50$ cm/s.  Rapid cadence and high signal-to-noise, high dispersion spectroscopy allows for better sampling and modeling of fluctuations caused by the contaminating micro-telluric lines. With  high temporal sampling and high spectral resolution enabled by the 30-meter telescopes, the telluric lines can be sampled and modeled sufficiently to reduce the induced noise to a few cm\,sec$^{-1}$ \cite{halverson2016}.  Finally, with the shorter integration times, the determination of the observation midpoints are improved and, hence, the accuracy of the barycentric correction can be improved by a factor of 3 or more to less than $\sim1$ cm/sec \cite{halverson2016}.  
\begin{figure}[h]
\vspace{-0pt}
    \begin{minipage}[c]{0.45\textwidth}
        \includegraphics[width=\textwidth]{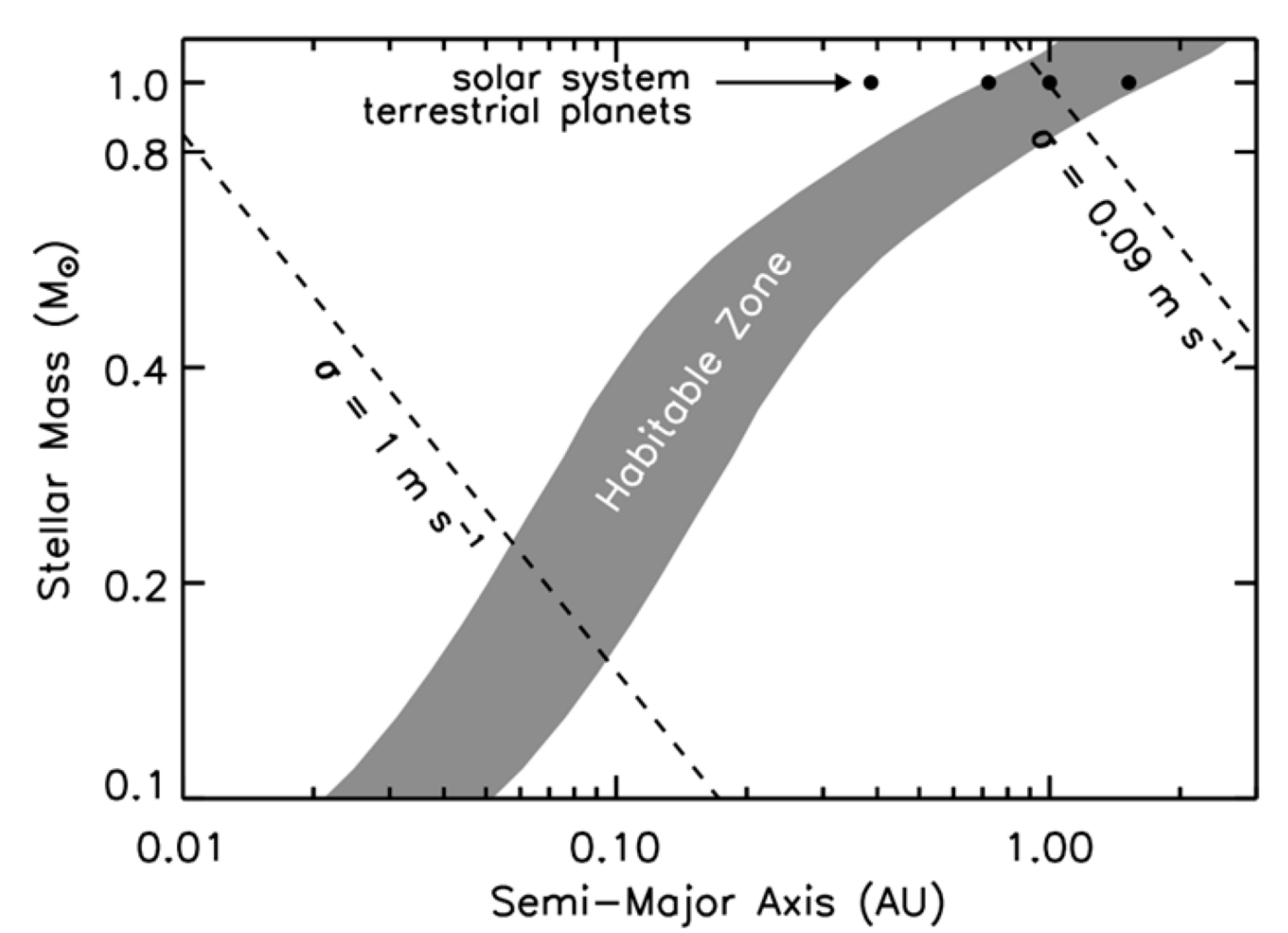}
    \end{minipage}\hfill
    \begin{minipage}[c]{0.55\textwidth}
        \vspace{-20pt}
        \caption{Radial velocity sensitivity requirements as a function of stellar mass and semimajor axis. The shaded region indicates the location of the traditional liquid water habitable zone \cite{selsis2007}. The dashed lines show where the radial velocity semi-amplitude of a 1 M$_\oplus$ planet would equal either 1 m/s or 0.09 m/s.  Adopted from NAS Exoplanet Strategy Report \cite{nas2018}.}
    \end{minipage}
    \vspace{-13pt}
\end{figure}

The proposed 30-m telescopes and the space-based probe EarthFinder can facilitate observations at a cadence that will enable mitigation of the stellar activity in a complementary manner to the ``jitter''-averaging approach done today with current facilities.  Additionally, enabling observations of fainter stars known to host long period transiting planets can allow studies of systems currently out of reach of today's facilities. The greater sensitivity of the 30-meter telescopes and the space-based platforms will allow for higher spectral resolution making it possible to measure and model the spectral line fluctuations at sufficient signal-to-noise to enable the disentangling of stellar activity radial velocity signatures, which are wavelength-dependent, from Keplerian orbit radial velocity signatures, which are wavelength-independent \cite{dumusque2018}.\\

\noindent\underline{\textbf{Recommendations:}}
Over the next decade, humanity has opportunity to address fundamental questions we have been asking for millenia: \textbf{Are we Alone?} In order to find and characterize Earth 2.0, mass and orbit measurements of Earth-sized planets in the habitable zone of their host stars are crucial.  Toward that end, we have three major recommendations:\\
\noindent 1. We endorse the findings and recommendations published in the National Academy reports on Exoplanet Science Strategy and Astrobiology Strategy for the Search for Life in the Universe. This white paper extends and complements the material presented therein.\\
\noindent 2. Specifically, the US should invest in the EPRV initiative as recommended by the National Academies of Sciences Exoplanet Science Strategy Report \cite{nas2018}.\\
\noindent 3. The US should invest in the instrumentation, telescope facilities (ground and space-based) and advanced tools for statistical modeling of stellar variability  necessary to obtain the radial velocity observations at the precision and quality sufficient to determine the masses and orbits of Earth-sized planets in the habitable zones of Sun-like stars.  The most compelling systems will be discovered in both hemispheres, and thus, facilities need to access the whole sky.

\pagebreak

\end{document}